\documentclass[final,5p,times]{elsarticle}
\usepackage{amssymb}
\usepackage{color}
\usepackage{amsmath}

\journal{Elsevier journal}

\newcommand{\blue}[0]{\textcolor{black}}

\usepackage[hang,small,bf]{caption}
\usepackage[subrefformat=parens]{subcaption}
\begin{document}

\begin{frontmatter}

\title{Gaussian Process Phase Interpolation for estimating the asymptotic phase of a limit cycle oscillator from time series data}

\author[Tokyo1]{Taichi Yamamoto}
\affiliation[Tokyo1]{organization={Graduate School of Frontier Sciences, The University of Tokyo},   
city={Kashiwa}, postcode={277-8561}, 
country={Japan}}
\author[TokyoTec1,TokyoTec2]{Hiroya Nakao}
\affiliation[TokyoTec1]{organization={Department of Systems and Control Engineering, Institute of Science Tokyo},  
            city={Tokyo},  postcode={152-8552},  
            country={Japan}}  
\affiliation[TokyoTec2]{organization={Research Center for Autonomous Systems Materialogy, Institute of Science Tokyo},  
            city={Yokohama},  postcode={226-8501},  
            country={Japan}}  
\author[Tokyo1,Tokyo2]{Ryota Kobayashi}
\ead{r-koba@k.u-tokyo.ac.jp}
\affiliation[Tokyo2]{organization={Mathematics and Informatics Center, The University of Tokyo},   
city={Tokyo}, postcode={113-8656}, 
country={Japan}}            

\begin{abstract}
Rhythmic activity commonly observed in biological systems, occurring from the cellular level to the organismic level, is typically modeled as limit cycle oscillators. 
Phase reduction theory serves as a useful analytical framework for elucidating the synchronization mechanism of these oscillators. Essentially, this theory describes the dynamics of a multi-dimensional nonlinear oscillator using a single variable called asymptotic phase. 
In order to understand and control the rhythmic phenomena in the real world, it is crucial to estimate the asymptotic phase from the observed data. 
In this study, we propose a new method, Gaussian Process Phase Interpolation (GPPI), for estimating the asymptotic phase from time series data. The GPPI method first evaluates the asymptotic phase on the limit cycle and subsequently estimates the asymptotic phase outside the limit cycle employing Gaussian process regression. Thanks to the high expressive power of Gaussian processes, the GPPI is capable of capturing a variety of functions.
Furthermore, it is easily applicable even when the dimension of the system increases.
The performance of the GPPI is tested by using simulation data from the Stuart-Landau oscillator and the Hodgkin-Huxley oscillator. The results demonstrate that the GPPI can accurately estimate the asymptotic phase even in the presence of high observation noise and strong nonlinearity. Additionally, the GPPI is demonstrated as an effective tool for data-driven phase control of a Hodgkin-Huxley oscillator. 
Thus, the proposed GPPI will facilitate the data-driven modeling of the limit cycle oscillators. 
\end{abstract}

\end{frontmatter}

\section{Introduction}
\label{introduction}

Rhythmic activity is a ubiquitous phenomenon observed in a wide range of biological systems. 
Examples include cortical networks in the brain~\citep{buzsaki2004,wang2010}, circadian rhythms in mammals~\citep{mohawk2011cell,yamaguchi2013mice}, the human heart and respiratory system~\citep{schafer1998,lotrivc2000,kralemann2013vivo}, and animal gait~\citep{collins1993coupled,kiehn2006locomotor,funato2016evaluation,kobayashi2016estimation}. Most mathematical models of rhythmic activity are based on limit cycle oscillators~\citep{strogatz2018nonlinear}. 
Phase reduction theory~\citep{kuramoto1984chemical,hoppensteadt2012weakly,winfree2001geometry,ermentrout2010mathematical,nakao2016phase,monga2019phase,kuramoto2019concept,ermentrout2019recent,leon2023analytical} is a valuable tool for analyzing the synchronization of limit cycle oscillators. This theory represents the state of a multi-dimensional limit cycle oscillator using a single variable, called the phase, and describes their dynamics in a reduced phase model. Theoretical studies based on the phase model have elucidated the key factors underlying synchronization phenomena, including the periodic external force, the coupling between oscillators, and the common inputs~\citep{nakao2016phase}. 

A fundamental challenge in the field of complex systems is to identify the mechanisms by which a real-world system achieves synchronization~\citep{pikovsky2001synchronization}. 
While these theoretical studies offer explanations for synchronization phenomena, they require the precise knowledge of the mathematical models. 
Consequently, previous studies have developed methods for identifying the phase model from real data (for a review, see \cite{stankovski2017coupling,rosenblum2023inferring}).

The most common approach for identifying the phase model from data is to estimate the phase sensitivity function (also known as the  infinitesimal phase response curve), which characterizes the linear response property of the oscillator phase~\citep{galan2005efficient,tsubo2007,ota_aoyagi2009weighted,ota_aonishi2009map,nakae2010bayesian,stankovski2012inference,imai2017robust,cestnik2018inferring,matsuki2024network}. 
However, when a limit cycle oscillator is subjected to strong perturbations, such as a strong impulse input, the approximation using the phase sensitivity function may be inaccurate. 
Therefore, it is essential to obtain the asymptotic phase distant from the limit cycle in order to understand the synchronization property and to control the phase of the limit cycle oscillator subjected to strong perturbations. Recently, \citet{namura2022estimating} proposed a method to obtain the asymptotic phase in a data-driven manner. However, it is still challenging to estimate the asymptotic phase of high-dimensional systems. 

In this study, we propose the Gaussian Process Phase Interpolation (GPPI) method, which estimates the asymptotic phase from time series obtained from a limit cycle oscillator without assuming the mathematical model of the system. The GPPI method can be readily applied to systems with more than three dimensions, as the same algorithm is used regardless of the dimensionality of the dynamical system. 
To validate the GPPI, we apply it to two limit cycle oscillators: the Stuart-Landau oscillator and the Hodgkin-Huxley oscillator. 
Furthermore, we apply the GPPI to control the phase of the Hodgkin-Huxley oscillator by impulse inputs, which demonstrates the effectiveness of the GPPI for data-driven control. 

This paper is organized as follows: Sec.~\ref{sec_definition} outlines the phase reduction theory and the asymptotic phase. Sec.~\ref{sec_method} describes the proposed method, the GPPI, for estimating the asymptotic phase. In Sec.~\ref{sec_result}, the GPPI is validated using synthetic data obtained from the two limit cycle oscillators. In Sec.~\ref{sec_control}, we demonstrate data-driven control of a Hodgkin-Huxley oscillator by impulse input using the estimated asymptotic phase. Finally, Sec.~\ref{sec_discussion} summarizes the results and discusses the conclusions.

\begin{figure*}[htb]
  \centering
       \includegraphics[width=18cm]{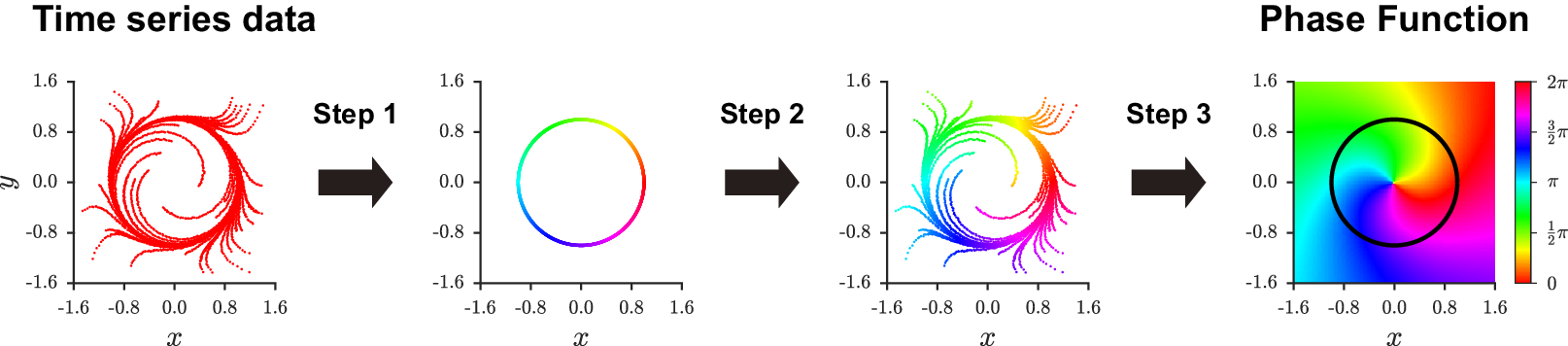}
  \caption{Procedure of the proposed method, Gaussian Process Phase Interpolation (GPPI). First, we determine the value of the phase function on the limit cycle (Step 1, Sec.~\ref{sec_Step1}). Then, we obtain the phase value of time series data (Step 2, Sec.~\ref{sec_Step2}). Finally, we estimate the global phase function by using Gaussian process regression (Step 3, Sec.~\ref{sec_Step3}).
}
  \label{fig:method}
\end{figure*}

\section{Phase reduction of limit cycle oscillators}
\label{sec_definition}
\newcommand{\R}{\mathbb{R}}
\newcommand{\x}{\mathbf{x}}
\newcommand{\X}{\mathbf{X_0}}

We consider a limit cycle oscillator described by
\begin{equation}
	\frac{d}{dt}\x(t) = \mathbf{F}(\x(t)), 
	\label{eq_dynamics}
\end{equation}
where $\x(t)\in \R^d$ is the state vector of the system at time $t$ and $\mathbf{F}:~\R^d\to\R^d$ is a smooth vector field representing dynamics. 
We assume that the system has a stable limit cycle orbit $\X(t)$ with the natural period and frequency, $T$ and $\omega= 2\pi/T$, respectively, which satisfies the condition: $\X(t+T) = \X(t)$. 

For any point in the basin of the limit cycle, an asymptotic phase can be determined~\citep{winfree2001geometry,nakao2016phase}. The basin $\mathcal{B}$ is defined as the set of initial conditions that converge to the limit cycle. 
In the following, the asymptotic phase is simply written as the `phase'. 

First, we assign the phase $\theta \in [0,2\pi)$ on the limit cycle, where the phase $\theta= 0$ and $2 \pi$ are considered to be identical. 
We choose a point on the limit cycle, which is set to be the phase origin, $\theta=0$, and we define the phase to increase at a constant rate $\omega$, i.e., $\theta(t) = \omega t$ (mod $2\pi$). 
As a result, the phase of the state $\X(t)$ on the limit cycle follows 
\begin{equation}
	\frac{d}{dt}\Theta(\X(t)) = \omega, 
	\label{eq_phasecycle}
\end{equation} 
where the phase function $\Theta(\x)$ gives the phase $\theta$ of the state $\x$ on the limit cycle. 
Moreover, the phase function can be extended so that Eq. (\ref{eq_phasecycle}) holds for any orbit $\x(t)$ in the basin of the limit cycle $\mathcal{B}$.
\begin{equation}
	\label{eq_phasediff}
	\frac{d}{dt}\Theta(\x(t)) = \omega.  
\end{equation}
Here, the domain of the phase function is extended as $\Theta:~\mathcal{B}\to[0,2\pi)$. 
The above procedure reduces the $d$-dimensional differential equation (\ref{eq_dynamics}) to a simple one-dimensional equation (\ref{eq_phasediff}) for any $\x(t) \in \mathcal{B}$. By integrating Eq.(\ref{eq_phasediff}), we obtain 
\begin{equation}
	\label{eq_phasestep}
	\Theta(\x(t+\tau)) - \Theta(\x(t)) = \omega\tau  \ \  
    ({\rm mod}\ 2\pi).  
\end{equation}

\renewcommand{\X}{\mathcal X_0}
Next, we introduce the phase response function (PRF, also known as the phase response or resetting curve, PRC)~\citep{kuramoto1984chemical,winfree2001geometry,brown2004phase,ermentrout2010mathematical}, which is essential for the analysis and control of synchronization phenomena. 
The PRF is a function that describes the effect of an impulse perturbation applied to an oscillator in phase $\theta$ and is defined as
\begin{equation}
    g(\theta,   {\boldsymbol  \varsigma}  ) = \Theta(\X(\theta)+ {\boldsymbol  \varsigma} ) - \Theta(\X(\theta)) = \Theta(\X(\theta)+ {\boldsymbol  \varsigma} ) - \theta,   
\end{equation}
where $ {\boldsymbol  \varsigma} \in \R^d$ denotes the intensity and direction of the impulse applied to the system (\ref{eq_dynamics})~\citep{winfree2001geometry,nakao2005synchrony,ermentrout2010mathematical,nakao2016phase}, and 
$\mathcal X_0(\theta)$ denotes the point with phase $\theta$ on the limit cycle. 
If the impulse intensity $\| \boldsymbol \varsigma\|$ is sufficiently small, the PRF can be approximated as 
\begin{equation}
    g(\theta, {\boldsymbol  \varsigma} ) = \mathbf{Z}(\theta)\cdot  {\boldsymbol  \varsigma} +O(\|\varsigma\|^2)
\end{equation}
where
\begin{equation}
    \mathbf{Z}(\theta) = \nabla\Theta(\X(\theta))
\end{equation}
is called the phase sensitivity function (PSF, also known as the infinitesimal phase resetting curve, iPRC). The PSF characterizes the linear response to a given weak perturbation for the state of phase $\theta$.

\section{Proposed method}
\label{sec_method}
We propose a method, Gaussian Process Phase Interpolation (GPPI), for estimating the phase function $\Theta(\x)$ from time series data of multiple orbits converging to the limit cycle. 
The main idea is to obtain the phase value of time series data using the phase equation \eqref{eq_phasestep} and interpolate them by Gaussian process regression. 
The procedure can be divided into the following three steps (Fig. \ref{fig:method}):   
\begin{itemize}    
    \item[1.] Determine the phase value on the limit cycle. 
    \item[2.] Obtain the phase value of the time series data.
    \item[3.] Interpolate the phase function using Gaussian process regression.
\end{itemize}
\newcommand{\homega}{\hat\omega}

\subsection{Step 1: Determine the phase value on the limit cycle}  
\label{sec_Step1}

In the first step, we obtain the value of the phase function on the limit cycle. 
Let us assume that a sufficiently long time series, $\x_0(t)$, is obtained from the limit cycle oscillator~\eqref{eq_dynamics} and that at time $t=0$, the orbit has converged sufficiently to the limit cycle.  
We first estimate the natural frequency $\omega$. 
Let us consider a Poincar\'{e} section that intersects the limit cycle once.  
The period and frequency of the limit cycle can be estimated using the following formulae: $\hat T=(s_{n+1}-s_1)/n$ and by $\homega = 2\pi / \hat T$, where $s_1$ and $s_{n+1}$ are the times of the first and $(n+1)$-th passage through the Poincar\'{e} section, respectively. 
In order to determine the phase, a point $\x_0(t_0)$ ($t_0 \geq 0$) on the time series is taken as the origin of the phase, i.e., $\Theta(\x_0(t_0))=0$. Subsequently, the phase on the limit cycle can be obtained using the phase equation (\ref{eq_phasestep}): $\Theta(\x_0(t)) = \homega(t-t_0)$ for any $t \geq 0$. 

\subsection{Step 2: Obtain the phase value of the time series data.} 
\label{sec_Step2}

In the second step, we obtain the phase value on each data point outside the limit cycle. 
Let us assume that we have collected $n$ time series, represented by $\{ \x_1(t), \cdots, \x_n(t) \}$, which converge to the limit cycle from its basin. 
It should be noted that such time series can be collected, for example, by applying perturbations to a limit cycle oscillator and observing the convergence to the limit cycle. 

We describe the procedure for obtaining the phase value on a data point that is outside the limit cycle. Let us consider a time series $\x_i(t)$ whose last point $\x_i(t_{{\rm end} })$ is close enough to the limit cycle. 
We can calculate the phase of the last point, $\Theta(\x_i(t_{{\rm end} }))$, by applying 
the linear interpolation of the data points on the limit cycle, $\Theta(\x_0(t))$, obtained in Step 1. 
Then, the phase at any time $t$ on the time series can be calculated from the last point using the phase equation \eqref{eq_phasestep} as
\begin{equation}
    \Theta(\x_i(t)) = \Theta(\x_i(t_{{\rm end}})) + \homega( t- t_{{\rm end}} ).
\end{equation}
In this way, the phase of all points on the time series $\x_i(t)$ are obtained. Similarly, the phase of each data point can be obtained by repeating this procedure for all time series ($i= 1, 2, \cdots, n$). 
In this study, the entire time series is discarded if the last point is not sufficiently close to the limit cycle.

\subsection{Step 3: Interpolate the phase function using Gaussian process regression}
\label{sec_Step3}  

In the third step, we estimate the global phase function $\Theta(\x)$ by employing Gaussian process regression~\citep{RasmussenW06,Schulz2018}. 
The proposed method does not directly estimate the phase function, $\Theta(\x)$, but rather estimates two real-valued functions 
\begin{align}
    s(\x) &:= \sin{\Theta(\x)}, \\
    c(\x) &:= \cos{\Theta(\x)}, 
\end{align}
and obtains the phase function from these functions. 

First, we generate two datasets $\{(\x_i,s(\x_i))\}$ and $\{(\x_i,c(\x_i))\}$ $(i=1, 2, \cdots, N)$ from the data $\{(\x_i,\Theta(\x_i))\}$ obtained in step 2 (Sec.~\ref{sec_Step2}). 
Next, we estimate the two functions $(s(\x), c(\x))$ from the datasets using Gaussian process regression. 
Finally, the phase function is estimated as 
\begin{equation}
    \hat\Theta(\x)=\arctan(\hat s(\x)/\hat c(\x)),  
    \label{eq_arctan}
\end{equation}
where $\hat s(\x)$ and $\hat c(\x)$ are the estimates of Gaussian process regression. 

\blue{
Here we describe the procedure for estimating the function $s(\x)$ using Gaussian process regression. It should be noted that the same procedure is applied to estimate $c(\x)$. 
The estimate $\hat s(\x)$ is obtained by calculating a Bayesian estimator, which is the conditional expectation given the observed data, $\hat s(\x)= E[ s(\x) | s(\x_1), s(\x_2), \cdots,  s(\x_N) \}$.  
Assuming that the function follows a Gaussian process, the conditional expectation is written as follows:
}
\renewcommand{\k}{\mathbf{k}}
\newcommand{\y}{\mathbf{y}}

\begin{align}
    \blue{\hat s(\x)} =         \k(\x)^\top{(K+\sigma^2I)}^{-1}\mathbf{s}, 
\end{align}
where 
\begin{align}
    K&\in\mathbb R^{N\times N},&
    K_{ij} &= k(\x_i,\x_j), \notag\\
    \k(\x)&\in\mathbb R^N,&
    \k(\x) &= ( k(\x_1,\x ), k(\x_2,\x ), \cdots , k(\x_N,\x ) )^\top, \notag\\  
    \blue{\mathbf{s}}&\in\mathbb R^N,&
    \blue{\mathbf{s}} &=
    \blue{(s(\x_1), s(\x_2), \cdots , s(\x_N))^\top}, \notag
\end{align}
$I \in \mathbb R^{N\times N}$ is the identity matrix, 
and $\sigma^2$ is the variance of the observation noise, respectively. 
The kernel function $k(\x_i,\x_j)$ represents the correlation \blue{between the values $s(\x_i)$ and $s(\x_j)$}. 
\blue{
Here, we adopt Mat\'{e}rn kernel with $\nu= 5/2$ for the kernel function:}  
\renewcommand{\top}{\mathsf{T}}
\begin{align}\blue{
    k(\x_i,\x_j) = \sigma_f^2
    \left(1+\frac{\sqrt5r}{\sigma_l}+\frac{5r^2}{3\sigma_l^2}\right)
    \exp{\left(-\frac{\sqrt5r}{\sigma_l}\right)},
    }
\end{align}
where $r= \sqrt{(\x_i-\x_j)^\top(\x_i-\x_j)}$ is the distance between the data points, and $\sigma_f$ and $\sigma_l$ are hyperparameters that determine the height and width of the kernel function, respectively. 
The hyperparameters are determined by maximizing the log-likelihood function given by 
\begin{multline}
    \log{p(\blue{\mathbf{s}} \mid \blue{\sigma_f, \sigma_l})}\\
    = -\frac12\blue{\mathbf{s}}^\top{(K+\sigma^2I)}^{-1}\blue{\mathbf{s}}
    -\frac12\log{\det{(K+\sigma^2I)}}-\frac N2\log{2\pi}. 
\end{multline}
It should be noted that the noise variance is set to $\sigma^2=0.01$. 
In this study, we used Gaussian process regression algorithm  implemented in MATLAB Statistical and Machine Learning Toolbox~\citep{MatlabStatML}. 
\blue{The source code of GPPI will be available on GitHub (https://github.com/TaichiY496/GPPI).}  

\section{Test of the proposed method}
\label{sec_result}
In this section, we test the validity of the proposed method, GPPI, by applying it to two limit cycle oscillators, the Stuart-Landau oscillator and the Hodgkin-Huxley oscillator.

\subsection{Evaluation methods}
\blue{
We test the proposed method by calculating the true phase function and comparing it with the estimation results. 
The true phase function of the Stuart-Landau oscillator is obtained from the analytical result~\citep{nakao2016phase}. 
For the Hodgkin-Huxley oscillator, the true phase function is obtained through exhaustive numerical simulation based on the mathematical model \eqref{eq_dynamics} in the following manner.} 
First, we obtain the phase function on the limit cycle in the same way as in the proposed method (Step 1, Sec.~\ref{sec_Step1}). Then, for each state $\x$ in a region in the basin, we simulate the dynamical system (\ref{eq_dynamics}) from $\x$ until the orbit converges to the limit cycle. Finally, the phase function $\Theta(\x)$ is obtained using the phase equation \eqref{eq_phasestep} in the same way as in the proposed method (Step 2, Sec.\ref{sec_Step2}). 
In this study, we use a two-dimensional grid in the state space, and calculate the phase function at all grid points. It is noteworthy that we use noiseless simulation to calculate the true phase function.

Moreover, we test the proposed method by comparing the normalized phase response function (nPRF) with its estimate.  
The nPRF for the $x$ direction is defined as $G_{x, \kappa}(\theta) := g(\theta, \kappa \mathbf{e}_x) / \kappa$, where $\mathbf{e}_x$ is the unit vector in the $x$ direction and $\kappa$ is the intensity of the impulse. 
The nPRF can be calculated from the estimate of the phase function $\hat\Theta$ as follows: 
\begin{equation}
    \hat G_{x,\kappa}(\theta) :=
    \frac{\hat\Theta(\X(\theta)+ \kappa \mathbf{e}_x) - \Theta(\X(\theta))}{\kappa},
\end{equation}
where $\X(\theta)$ denotes a state of the phase $\theta$ on the limit cycle.   

The estimation accuracy of the nPRFs is evaluated using the coefficient of determination defined as follows:
\begin{equation}
    R_Y^2 = 1 - \frac
    {\sum_i{\left(Y(\theta_i)-\hat Y(\theta_i)\right)^2}}
    {\sum_i{\left(Y(\theta_i)-\bar Y\right)^2}},
\end{equation}
where $Y$ and $\hat Y$ are the true and estimated values of the nPRF, respectively, and $\bar Y$ is the average of $Y$ over all the phase $\theta_i$. The closer the coefficient of determination is to 1, the higher the estimation accuracy is.

\subsection{Stuart-Landau oscillator}
\label{sec:SL_model}

\begin{figure*}[htb]
  \centering
       \includegraphics[width=17cm]{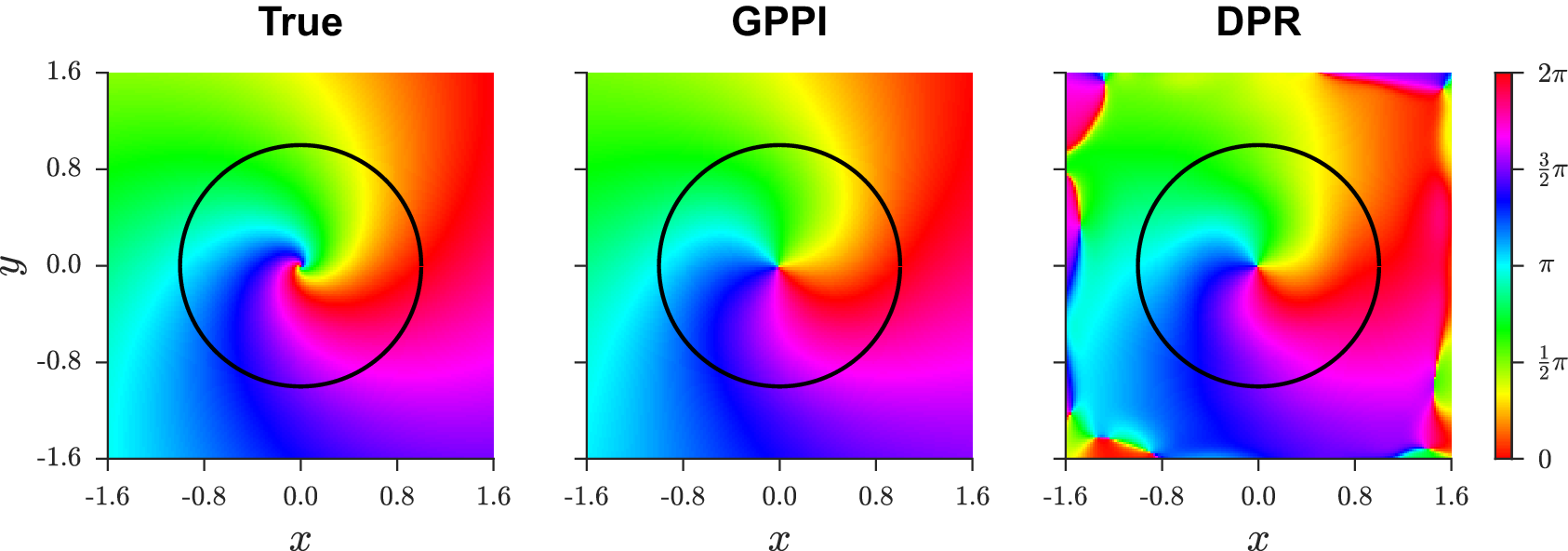}
  \caption{ 
  Test of the proposed method using the Stuart-Landau (SL) oscillator. The true phase function (True, left) was compared with its estimate based on the proposed method (GPPI, center) and an existing method (DPR, right) from time series data. The black circles represent the limit cycle orbit of the SL oscillator.  
}
  \label{fig:SLphase}
\end{figure*}

\begin{figure*}[htb]
  \centering
       \includegraphics[width=17cm]{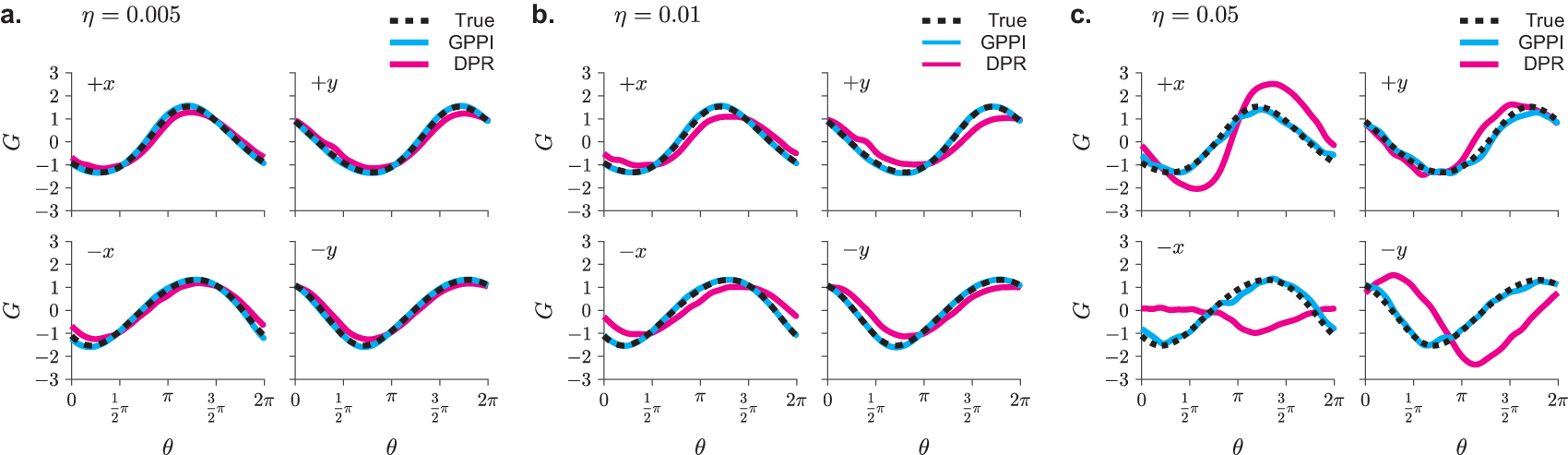}
  \caption{Effect of the observation noise on the estimation of the normalized phase response functions (nPRFs). The nPRFs calculated from the true phase function (True, black dashed line) were compared with the nPRFs estimated by using the proposed method (GPPI, cyan) and an existing method (DPR, magenta). Three levels of the observation noise $\eta$ are examined: a. $\eta= 0.005$, b. $\eta= 0.01$, and c. $\eta= 0.05$. The impulse intensity was set to $0.2$ for each direction. Note that the true phase function is calculated without noise. 
}
  \label{fig:SLprf}
\end{figure*}

We evaluate the proposed method using the Stuart-Landau (SL) oscillator and compare it with an existing method, which we refer to as the Derivative Phase Regression method (DPR method)~\citep{namura2022estimating}. We also examine the robustness of these methods against observation noise. 

The SL oscillator is described as follows:		
\begin{align}
    \dot x &= x - \alpha y - (x - \beta y)(x^2+y^2),		\label{eq:SL1}		\\
    \dot y &= \alpha x + y - (\beta x + y)(x^2+y^2),	\label{eq:SL2}
\end{align} 
where $x, y$ are the state variables and $\alpha=2, \beta=1$ are parameters. The numerical simulation was conducted using the fourth-order Runge$-$Kutta method with a time step of $\Delta t=0.005$. 

We generate synthetic data for validation as follows: an orbit is simulated based on Eqs.~(\ref{eq:SL1}) and (\ref{eq:SL2}) for the time length of $2.5$ from an initial state randomly sampled from the uniform distribution of ${[-1.6, 1.6]}^2$. It should be noted that the period of the limit cycle is $2 \pi$. The time series data $\{ x(t), y(t) \}$ is collected if the final state is sufficiently close to the limit cycle. This procedure is repeated until we obtain $n= 100$ time series. 
Finally, observation noise is added to the time series for testing the robustness against noise. 
The observation noise is given by independent Gaussian random variables of mean $0$ and standard deviation of $\eta=0.005$.

We first estimate the natural frequency (Step1, Sec.~\ref{sec_Step1}). To reduce the impact of observation noise, a moving average with a window width of $0.07$ was calculated from the time series data.
The natural frequency is estimated as $\homega= 1.000$, which is in agreement with the theoretical value of $\omega=1.0$ derived from the model. The estimated frequency $\homega= 1.000$ is also used in the DPR method. 
Next, we estimate the phase function $\Theta$ from the time series using the proposed method, the GPPI, as well as the DPR. 
To evaluate the performance of the proposed method for a small dataset, 1,000 data points are sampled with a coarse interval $\Delta t = 0.25$, and are used as a training data set for Gaussian process regression. 
In contrast, the DPR uses all simulation data, 50,000 points with sampling interval of $\Delta t = 0.005$. The other parameters of the DPR were the same as those used in \citet{namura2022estimating} (see Appendix A for details).

Figure \ref{fig:SLphase} compares the true phase function with its estimate by the proposed GPPI method and the DPR method from the time series. 
While the both methods accurately estimate the phase function near the limit cycle, they cannot estimate the phase function around the origin $(x,y)=(0,0)$ due to the lack of data. 
In particular, in the boundary region, the GPPI provides a more accurate estimate of the phase function than the DPR (see Fig.~\ref{fig:Append}a in Appendix B).  
One potential explanation for the decline in the estimation performance of the DPR is the use of a polynomial function, which tends to diverge the absolute value of the function in the boundary region. Instead of the polynomial function, the GPPI employs Gaussian process regression for estimating the phase function. 
Overall, the proposed GPPI method provides an accurate estimation of the phase function in the global region with smaller data compared to the DPR method (1,000 vs 50,000 data points). 

We also compared the normalized phase response function (nPRF) calculated from the estimated phase functions with the true values. Figure~\ref{fig:SLprf}a shows the true and estimated nPRFs obtained by applying impulses in four directions $(+x,-x,+y,-y)$ at an intensity of $0.2$.
Both methods yield accurate estimates of the phase response. However, the present GPPI method outperforms to the DPR method, with a higher coefficient of determination averaged over the four directions for the GPPI and the DPR ($0.998$ and $0.952$). 
It should be noted that the accuracy of the DPR is worse than the results in \citet{namura2022estimating} (Figure 3 and Table 1). This discrepancy may be attributed to the small number of time series used for the estimation.

Furthermore, we examine the robustness of the methods against observation noise by generating new data sets with increasing noise levels, specifically, $\eta=0.01, 0.05$. The estimation procedure was repeated for both methods using the new data, including the estimation of the natural frequency. 
Figures~\ref{fig:SLprf}b and c show the estimation results for nPRFs. As the noise strength increases, the estimation error of the DPR method also increases. The DPR is unable to estimate the phase response from noisy time series ($\eta=0.05$). 
In contrast, the GPPI is capable of estimating the phase response even in the presence of the substantial observation noise ($\eta=0.05$).  
Table~\ref{tab_SLprf} shows a comparison of the performance of the GPPI with that of the DPR based on the average of the  coefficient of determination $R_G^2$. 
These results demonstrate that the proposed GPPI method can estimate the phase function with less data than the DPR method and it is also robust against noise.

\begin{table}[tbp]
    \caption{Average coefficients of determination of the nPRFs of the SL oscillator. For three noise levels $\eta$ and two methods (the GPPI and the DPR), we calculate the average of the coefficient of determination, $R_G^2$, for impulses applied in four directions ($x+, x-, y+, y-$). The impulses are the same as those used in Fig.~\ref{fig:SLprf}. The higher score is shown in 
    bold. }   
    \centering
    \begin{tabular}{c|ccc}
    \hline\hline
       Method  & $\eta=0.005$  & $\eta=0.01$ & $\eta=0.05$ \\
       \hline\hline
       GPPI  & \textbf{0.998} & \blue{\textbf{0.999}} & \blue{\textbf{0.975}} \\
       DPR  & 0.952 & 0.831 & -0.409 \\       
    \hline\hline
    \end{tabular}
    \label{tab_SLprf}
\end{table}

\subsection{Hodgkin-Huxley oscillator}
\label{sec:HH_model}

As a second example, we test the proposed method using the Hodgkin-Huxley model (HH model), which is a four-dimensional dynamical system. The HH model has a higher dimension than the SL oscillator and also exhibits stronger nonlinearity than the SL oscillator.

The HH model is one of the most important neuron models in neuroscience, which has been used to describe neural systems mathematically~\citep{Dayan2005,Kobayashi2013}. This model is written as a differential equation of four variables, namely, $V, m, h$, and $n$, as follows~\citep{hodgkin1952quantitative}:

\newcommand{\g}[1]{\bar g_\text{#1}}
\newcommand{\vv}[1]{V_\text{#1}}
\begin{align}    
    \dot V =& \frac1C \Bigl( -\g{Na}(V-\vv{Na})m^3h-\g{K}(V-\vv{K})n^4 \nonumber\\ & -\g{L}(V-\vv{L}) + I_b \Bigr),  \label{eq_HHV}    \\    
    \dot m =& \alpha_m(V)(1-m) - \beta_m(V)m,	  \label{eq_HHm}    \\
    \dot h =& \alpha_h(V)(1-h) - \beta_h(V)h,	  \label{eq_HHh}    \\    
    \dot n =& \alpha_n(V)(1-n) - \beta_n(V)n,     \label{eq_HHn}
\end{align}
where the functions $\alpha_{m,h,n}, \beta_{m,h,n}$ are written as
\begin{align*}
    \alpha_m(V) &= \frac{0.1(V+40)}{1-e^{-(V+40)/10}},
    &\beta_m(V) &= 4e^{-(V+65)/18},\\
    \alpha_h(V) &= 0.07e^{-(V+65)/{20}},
    &\beta_h(V) &= \frac{1}{1+e^{-(V+35)/10}},\\
    \alpha_n(V) &= \frac{0.01(V+55)}{1-e^{-(V+55)/10}},
    &\beta_n(V) &= 0.125e^{-(V+65)/80},  
\end{align*}
and the parameters are set as follows:  $C= 1$ [$\mu {\rm F/cm^2}$],  $\g{Na}  = 120$ [${\rm mS/cm^2}$],   $\g{K}= 36$ [${\rm mS/cm^2}$],   $\g{L}= 0.3$ [${\rm mS/cm^2}$], $\vv{Na} = 50$ [mV],  $\vv{K} = -77$ [mV], $\vv{L} = -54.4$ [mV],  and $I_b= 10$ [$\mu {\rm A/cm^2}$].  
The numerical simulation was conducted using the fourth-order Runge$-$Kutta (RK4) method with a time step $\Delta t=0.01$ [ms].

\begin{figure}[tb]
  \centering
       \includegraphics[width=8.5cm]{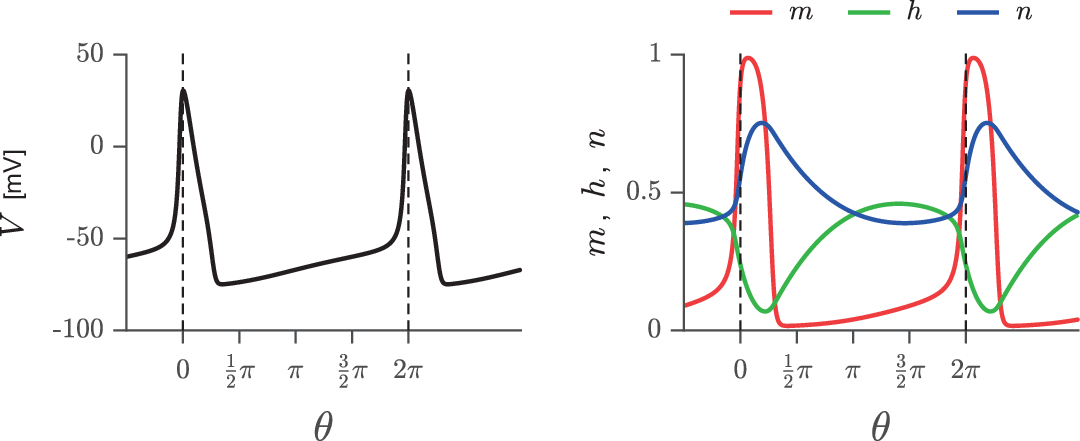}
  \caption{State variables of the Hodgkin-Huxley (HH) oscillator. Here, the origin of the phase ($\theta= 0$) is defined as the peak of the voltage $V$ (the dashed line). } 
  \label{fig:HH_oscillator}
\end{figure}

A limit cycle that corresponds to periodic firing can be obtained from the HH model by setting the input current $I_b$ above a certain threshold~\citep{keener2009}. In this study, we focus on the oscillatory regime, where the HH model has a limit cycle.  
Figure~\ref{fig:HH_oscillator} illustrates a single period of the limit cycle obtained from the HH model. 
Following the literature~\citep{osinga2010continuation}, the origin of the phase is defined as the peak of the voltage $V$. 
The period and the natural frequency are estimated from a time series of 1,000 periods, which results in $\hat T=14.6$ [ms] and $\homega=0. 429$ [rad/ms]. It should be noted that these values are also used in the calculation of the true phase function, as we do not consider the observation noise in this example.

Here, we consider two strategies for collecting the time series data used to estimate the phase function. 
The first strategy is to simulate the system from the initial state, which is randomly sampled from a uniform distribution of a hyperrectangle, as in the case of the SL oscillator. For the HH oscillator, we employ a hyperrectangle formed by $V \in [-100, 50], m \in [0,1], h \in [0, 0.6], n \in [0.3, 0.8]$. 
We refer to this strategy as uniform sampling. The uniform sampling strategy is able to collect data over a global region of the state space. However, as the dimension of the state space increases, the data points become increasingly sparse, thereby rendering the estimation of the phase function more difficult. 
The second strategy is to simulate the system from the initial state generated by adding a small perturbation to the state on the limit cycle. For the perturbation of the HH oscillator, we employ a uniform random number in the range $[-\sigma_x/10, \sigma_x/10]$, where $\sigma_x$ is the standard deviation of the variable $x \in \{V, m, h, n \}$ along the limit cycle. 
We refer to this strategy as vicinity sampling. While the vicinity sampling is an effective strategy for accurately estimating the phase function of regions in proximity to the limit cycle, it is difficult to apply this strategy for estimating the phase function of regions distant from the limit cycle. 
We generate the orbits of time length 100 [ms] from 100 initial states obtained for each sampling strategy by simulating the HH model. Subsequently, the time series of each state variable is collected. It should be noted that, in uniform sampling, several time series that do not converge to the limit cycle are excluded from the analysis. 

The number of data points thus prepared is approximately $10^6$. 
Given the high computational cost associated with Gaussian process regression, it is necessary to reduce the number of data points to a range of $10^3$ to $10^4$. Accordingly, the training data used for Gaussian process regression is extracted from the data set in the following manner. 
First, we extract the initial part of the time series, which is relatively distant from the limit cycle. 
Specifically, the initial 20 [ms] of the time series with a sampling interval of 1 [ms] are extracted in the uniform sampling, and the initial  6 [ms] of the time series with a sampling interval of 0.3 [ms] are extracted in the vicinity sampling.  
Furthermore, due to the rapid change in the state variables before and after a spike, the data density in the region corresponding to the spike tends to be relatively low. To ensure sufficient data collection in the spike region, we employ a smaller sampling interval. Specifically, the sampling interval is decreased to $1/10$ during the spike period, $t_f-4 \le t \le t_f+3$ [ns], where $t_f$ is the spike time defined as the time when the voltage $V$ exceeds 0 [mV]~\citep{kobayashi2016impact}. 
Finally, 100 points from the limit cycle are added to the training data to supplement the phase information.

\begin{figure*}[htb]
  \centering
       \includegraphics[width=17cm]{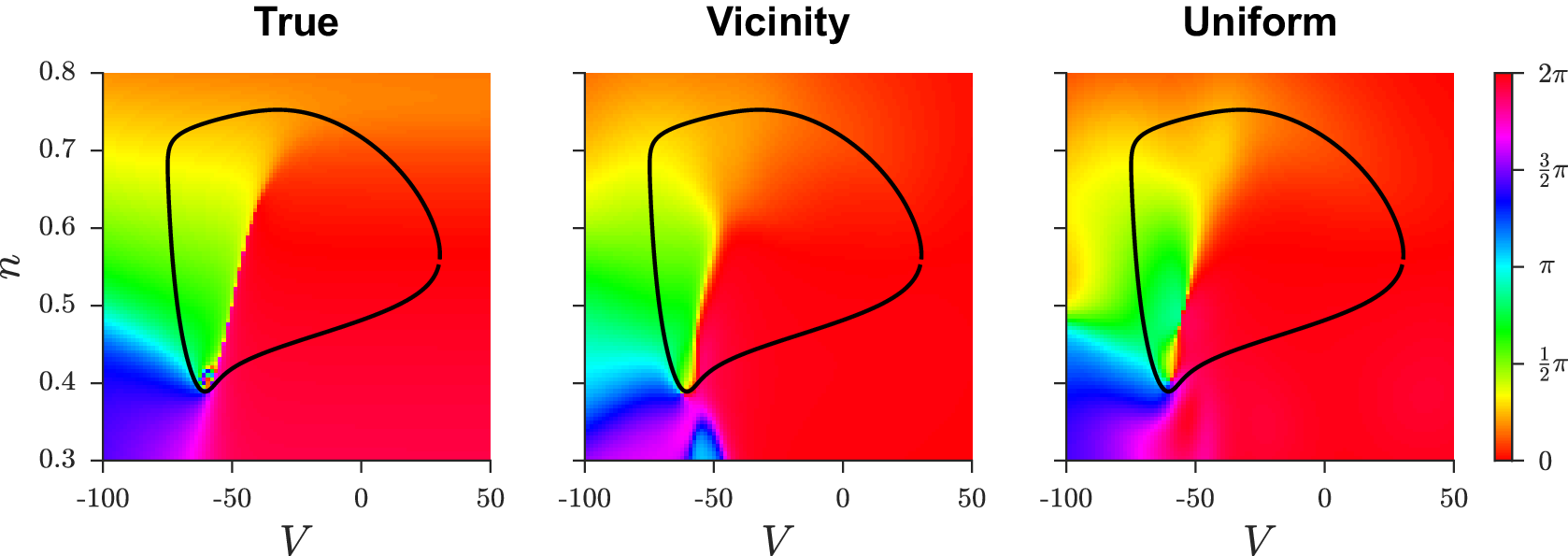}
  \caption{ 
  Test of the proposed method using the Hodgkin-Huxley (HH) oscillator. The true phase function (left) was compared with its estimate based on the proposed method from two datasets by using different sampling strategies: Vicinity sampling (center) and Uniform sampling (right). The black curve represents the limit cycle of the HH oscillator.  
}
  \label{fig:HHphase}
\end{figure*}
%
\begin{figure*}[htb]
  \centering
       \includegraphics[width=17cm]{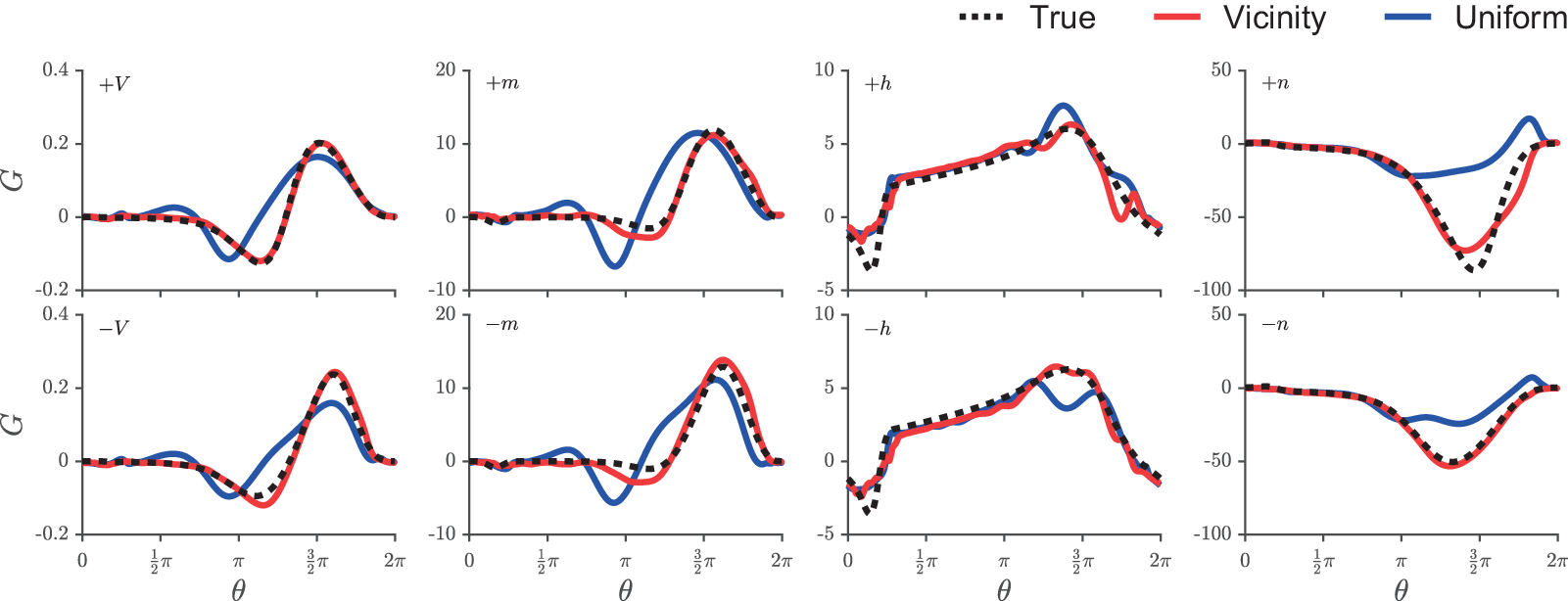}
  \caption{Estimation of the normalized phase response functions (nPRFs) from the HH oscillator. The nPRFs calculated from the true phase function (True, black dashed line) were compared with their estimate based on the proposed method with two sampling strategies: Vicinity sampling (red) and Uniform sampling (blue). The intensity of the impulse is set to $\sigma_i/20$, where $\sigma_i$ is the standard deviation of the variable $i=(V, h, m, n)$ over a period of the limit cycle.}  
  \label{fig:HHprf}
\end{figure*}

Figure \ref{fig:HHphase} compares the true phase function obtained from the HH model with its estimate by the proposed GPPI method using the two sampling strategies: uniform and vicinity sampling. The limit cycle of the HH model approximately lies on a surface that can be written as follows~\citep{fitzhugh1961,keener2009}:
\begin{align}
    \label{eq_HHsurf1}
    m &= m_\infty(V) := \alpha_m(V) / \{\alpha_m(V) + \beta_m(V)\},\\
    \label{eq_HHsurf2}
    h &= 0.8 - n.
\end{align} 
We therefore compare the phase function in the region $V\in[-100,50], n\in[0.3, 0.8]$ on this surface. 
Although the data obtained from the vicinity sampling is distributed over a small region in proximity to the limit cycle, this strategy still worked well for estimating the phase functions of a more global region (see Fig.~\ref{fig:Append}b in Appendix B). 
It should be noted that neither method is able to estimate the phase function in the vicinity of the equilibrium ($V\sim-60$ [mV], $n\sim0.5$), where the phase varies in a complex way~\citep{osinga2010continuation}.

\begin{table}[tbp]
    \caption{
    Average coefficients of determination of the nPRFs of the HH oscillator. For two sampling methods (Vicinity and Uniform), we calculate the average of the coefficient of determination, $R_G^2$, for impulses in two directions ($+, -$). The impulses are the same as those used in Fig.~\ref{fig:HHprf}. 
    The higher score is shown in bold.
    }
    \centering
    \begin{tabular}{c|cccc}
    \hline\hline
       Sampling  & $V$  & $m$ & $h$ & $n$ \\
       \hline\hline
       Vicinity & \textbf{0.965} & \textbf{0.874} & 0.930 & \textbf{0.932} \\
       Uniform  & 0.349 & 0.329 & \textbf{0.946} & -0.732 \\       
    \hline\hline
    \end{tabular}
    \label{tab:HHprf}
\end{table}

Next, we examine the estimation performance of normalized phase response functions (nPRFs). Figure~\ref{fig:HHprf} compares the estimate of the nPRF from the HH oscillators, which shows that the vicinity sampling provides better results than the uniform sampling. 
Additionally, the difference between the nPRF for the positive and negative impulses reflects the nonlinearity of the HH oscillator. While the HH oscillator has a strong nonlinearity in the $n$ direction, this nonlinearity can be captured by the vicinity sampling strategy. It should be noted that the intensity of the impulse used to calculate the nPRF is sufficiently low to fall within the range of the neighborhood sampling. 
Furthermore, we evaluate the estimation accuracy of the nPRF using the average of the coefficients of determination (Table~\ref{tab:HHprf}). 
As can be seen in Fig.~\ref{fig:HHprf}, the coefficients of determination of the vicinity sampling are much higher than those of the uniform sampling, except for the direction in $h$; both sampling methods provide accurate estimates of the nPRF in the $h$ direction.  

\section{Data-driven phase control using impulse input}\label{sec_control}

As an application of the estimated phase function, we examine the control of the phase of a limit cycle oscillator using an impulse input. 
As in the previous section, we consider the Hodgkin-Huxley (HH) oscillator, but we consider a scenario that is more feasible in the experiments. 
Let us suppose that a time series \blue{$\x(t) = (V(t),m(t),h(t),n(t))$} can be measured from the HH oscillator that is stimulated by an impulse input with the direction of $+V$. 
Our objective is to control the phase of the oscillator from the time series without assuming any knowledge of the mathematical model (Eqs.~\ref{eq_HHV}, \ref{eq_HHm}, \ref{eq_HHh}, and \ref{eq_HHn}). 
While the kinetic variables $(m(t),h(t),n(t))$ cannot be recorded in experiments, we can estimate them from the experimental data $V(t)$~\citep{kobayashi2011estimating,meliza2014estimating}.

\begin{figure}[thb]
  \centering
       \includegraphics[width=8.5cm]{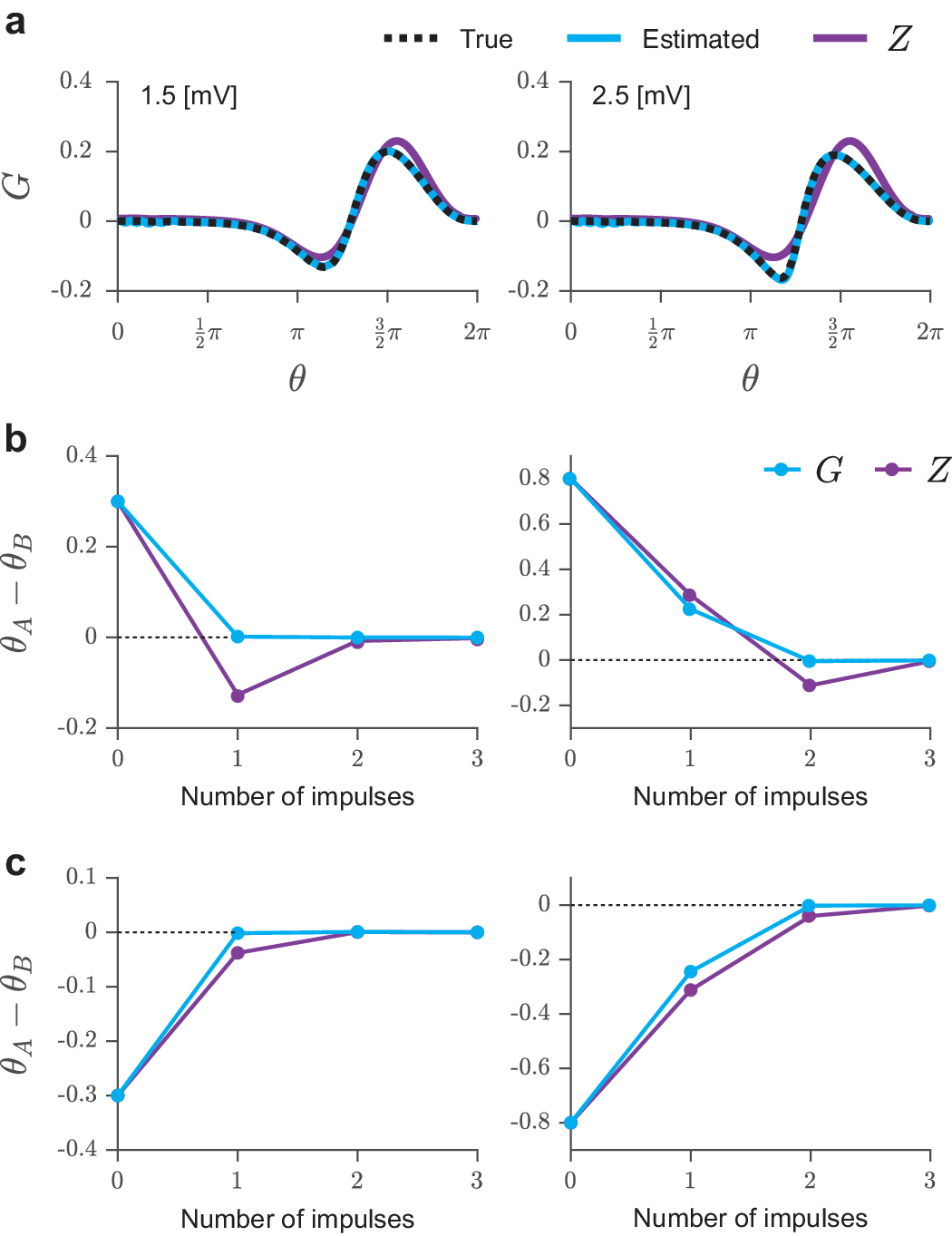}
  \caption{Data-driven phase control of the Hodgkin-Huxley (HH) oscillator. 
  (a) Estimation of the normalized phase response functions (nPRFs) of the HH oscillator. 
  The true nPRFs (dashed black) are compared with their estimates from the time series data (cyan) and their approximation based on the phase sensitivity function (purple). The impulse intensity was 1.5 [mV] (left) and 2.5 [mV] (right), respectively.   
  (b), (c) The phase control based on the estimated phase response ($G$) and that based on the phase sensitivity function ($Z$) are compared in four cases. The horizontal axis represents the number of impulse inputs, and the vertical axis represents the phase difference. 
  }
  \label{fig:DD_Control}
\end{figure}

A total of 90 time series are generated by applying impulses of intensity $\Delta V =1, 2,$ and $3$ [mV], respectively, to 30 points on the limit cycle of the HH oscillator. It should be noted that comparable time series data can be obtained in experiments where neurons are stimulated with repeated impulses.
We first estimate the phase function $\Theta(\x)$ using the proposed method (GPPI) from the time series data. The training data for the Gaussian process regression was prepared by sampling at 0.1 [ms] intervals from the initial 3 [ms] of the time series. We then calculate the normalized phase response functions (nPRFs) in the $+V$ direction from the estimated phase function. 
Figure~\ref{fig:DD_Control}a compares the true and estimated nPRFs for impulse intensities of $\Delta V = 1.5$ and $2.5$ [mV], as well as the phase sensitivity function $Z_V:= \frac{\partial\Theta}{\partial V}(\X(\theta))$. 
It should be noted that Figure~\ref{fig:DD_Control}a shows the prediction performance of the proposed method for the impulses that were not included in the training data set. 
In addition, the phase sensitivity function $Z_V$ was not estimated from the data, and it was obtained from the numerical simulation of the HH model with an impulse intensity of $\Delta V=0.1$ [mV].
For a strong impulse ($\Delta V =2.5$ [mV]), the true nPRF deviates from $Z_V$ due to nonlinearity of the oscillator, whereas the GPPI is able to estimate the phase response accurately from data.

Next, we synchronize the phases of two uncoupled HH oscillators (oscillator A and B) by using impulse inputs in the $+V$ direction to the oscillator A. 
The phases of the two oscillators ($\theta_A(t)$ and $\theta_B(t)$, respectively) are obtained from the observed data. The phase information is used to determine the intensity and timing of the impulse in order to align the phases of the two oscillators. 
Due to the limitation of the impulse intensity ($0 \leq \Delta V \leq 3$ [mV]), multiple impulses are required if the phase difference between the oscillators is too large. In such cases, the next impulse was added to the oscillator after sufficient time has elapsed to converge to the limit cycle. 
We examine a strategy for controlling the phase of the oscillator based on the phase response function (PRF): $g(\theta, \Delta V \mathbf{e}_V)$. The strategy is as follows: 
First, if the PRF indicates that it is possible to align the phases between the oscillators with a single impulse, we adopt the minimum strength of the impulse that achieves the complete synchronization. 
Second, if the PRF indicates that the phases cannot be aligned with a single impulse, we adopt the impulse that brings the phase difference closest to zero. In other words, if the phase of oscillator A is too far ahead (behind), the impulse that delays (advances) the phase the most is selected.

This control strategy based on the PRF obtained from the estimated phase function: $\hat g(\theta,\mathbf{e}_V\Delta V):= \hat\Theta(\X(\theta)+\mathbf{e}_V\Delta V) - \theta$, was applied to the HH oscillator. 
Figures~\ref{fig:DD_Control}b and c compare the performance of the phase control based on the estimated phase function with that based on the phase sensitivity function, where the PRF is approximated with the phase sensitivity function: $\hat g(\theta,\mathbf{e}_V\Delta V) \approx Z_V(\theta)\Delta V$. Again, the phase sensitivity function was not estimated from the data, and it was obtained from the HH model. 
Given that the maximum phase change due to an impulse is approximately 0.5, it is expected that a single impulse is sufficient when the initial phase difference is 0.3 (Fig.~\ref{fig:DD_Control}b and c: left). It can be also expected that two impulse inputs are required when the initial phase difference is 0.8 (Fig.~\ref{fig:DD_Control}b and c: right).
Figures~\ref{fig:DD_Control}b and c confirm that the phase of the HH oscillator can be controlled as expected from observation data alone, without the knowledge of the HH model.  
While the phase can be also controlled using the phase sensitivity function $Z_V$, an additional  impulse is required to achieve the complete synchronization. 
In particular, the method based on the phase sensitivity function becomes inefficient for delaying the phase (Fig.~\ref{fig:DD_Control}b). 
This is because the phase sensitivity function $Z_V$ overestimates the PRF when it takes negative values (Fig.~\ref{fig:DD_Control}a), yielding too strong input for the control. 
This result implies that the data-driven control can be achieved by using the proposed method (GPPI), and 
the GPPI method achieves superior performance to the method based on the phase sensitivity function obtained from the HH model.

\section{Discussion}
\label{sec_discussion}
In this study, we have proposed a method, Gaussian Process Phase Interpolation (GPPI), for estimating the asymptotic phase (i.e., the phase function) using time series data obtained from the limit cycle oscillator. This method is based on Gaussian process regression and is applicable even when the dimension of the dynamical system is increased. 
We have applied the method to the Stuart-Landau (SL) oscillator and the Hodgkin-Huxley (HH) oscillator. Our results demonstrate that the GPPI method can accurately estimate the phase function even in the presence of substantial observation noise and strong nonlinearity from relatively small dataset. 
Furthermore, we have demonstrated that the phase function estimated by the GPPI method is effective for data-driven phase control with impulse input.

As previously stated in the Introduction, the DPR method proposed in \cite{namura2022estimating} is an exiting method with the same objective as the proposed method. We have shown that the GPPI method can estimate the phase function with fewer data points (Fig.~\ref{fig:SLphase}) and is more robust to the observation noise than the DPR (Fig.~\ref{fig:SLprf} and Table~\ref{tab_SLprf}). 
The GPPI differs from the DPR mainly in two respects, which contribute to the improvement. The first difference is in the regression method. The GPPI employs Gaussian process regression, whereas the DPR is based on the polynomial regression. Gaussian processes have a higher expressive power than polynomial functions.
The second difference is in the object of the regression. The GPPI fits the functions of the asymptotic phase, i.e., $\sin{\Theta(\x)}, \cos{\Theta(\x)}$, whereas the DPR fits a differential equation that the phase satisfies (see~\ref{Append:DPR}). 
One advantage of the GPPI is that it can be applied to time series with a large sampling interval, because it does not require the derivative of the state variables.

The Phase AutoEncoder (PAE), recently proposed in \citet{yawata2024phase}, is comparable to the GPPI method in that it estimates the phase function from the time series. Moreover, it can be applied to limit cycle oscillators from high-dimensional dynamical systems. 
One advantage of the PAE is its capacity to learn the phase function from a large amount of data. 
\blue{
In contrast, 
the proposed method offers several advantages. Primarily, it is applicable to a moderate amount of data ($\sim$ 1,000 points), and the hyperparameters ($\sigma_f$ and $\sigma_l$) have intuitive interpretations, which allows for an understanding of the model behavior.}  

It has been shown that the asymptotic phase can also be interpreted as the argument of the Koopman eigenfunction~\citep{mauroy2012use,mauroy2013isostables,shirasaka2017phase}. 
The existing methods for estimating Koopman eigenfunctions, including Dynamic Mode Decomposition (DMD) and its extensions ~\citep{schmid2010dynamic,williams2015data,kutz2016dynamic} are applicable for estimating the asymptotic phase.  
However, there remain challenges for estimating the asymptotic phase, 
as an accurate estimate of the Koopman eigenvalue (i.e., the natural frequency) is required. 
In contrast, we separately estimated the accurate natural frequency by using the Poincar\'{e} section in the present method.

As an application of the phase function estimation, we have considered the problem of controlling the phase of an oscillator by impulse input. 
Specifically, we consider the HH oscillator and control the phase by using the pulse in the $+V$ direction, assuming that the time series $\{V(t), m(t), h(t), n(t) \}$ is available. 
While it is possible to stimulate the neuron using such an impulse in an experiment~\citep{galan2005efficient, tsubo2007}, the variables ($m, h$, and $n$), referred to as kinetic variables, cannot be measured directly. 
Nevertheless, it is possible to infer the kinetic variables from the measured data~\citep{kobayashi2011estimating,meliza2014estimating}. 
Consequently, the control of limit cycle oscillators in the real world, such as neurons and the human heart and respiratory system, using the proposed method would be one of the important directions of future research. 

A limitation of this study is the computational cost required for Gaussian process regression. A potential way to improve the estimation accuracy is to increase the amount of data. 
However, the computational complexity of Gaussian process regression is $O(N^3)$, where $N$ is the number of data points, rendering it an impractical method when data size increases. 
Indeed, the HH oscillator required approximately 10 times as many data points as the SL oscillator for the phase function estimation, which resulted in an actual computation time that was more than 100 times longer. 
A promising avenue for future research would be to improve the Gaussian process regression algorithm for more efficient phase function estimation. For example, we could use fast approximation algorithms for Gaussian process regression~\citep{liu2020gaussian}, including the FIC method~\citep{quinonero2005unifying}.    

For high-dimensional limit cycle oscillators, the data distribution in the state space may be non-uniform for the following reasons: 1) The dynamical system may possess multiple time constants, resulting in fast and slow parts of the limit cycle orbit, 2) The shape of the limit cycle may become complex in state space. 
It is essential to obtain the data points so that their distribution is close to uniform distribution when using the GPPI method. 
While the GPPI can accurately estimate the phase function of the HH oscillator (Fig. \ref{fig:HHphase} and \ref{fig:HHprf}), it requires proper data sampling. 
When we collect all the time series, the data tends to be concentrated in the vicinity of the limit cycle. Additionally, the data density is relatively low in regions where the state variable changes rapidly, such as the spike in the HH oscillator. 
It would be an interesting topic of future research to extend the GPPI so that it can be easily applied to a variety of limit cycle oscillators.

In this study, we have presented the Gaussian Process Phase Interpolation, a method for estimating the phase function of the limit cycle oscillator from time series data. We have demonstrated the utility of the proposed GPPI method by applying it to simulated limit cycle oscillators. Our future goal is to apply the GPPI to limit cycle oscillators in the real world and develop a methodology for controlling their synchronization dynamics. To this end, we will refine the phase estimation method so that it can handle more complex oscillatory systems.

\section*{Acknowledgements}
We thank Norihisa Namura for sharing us the code used in ~\cite{namura2022estimating}. We thank Hiroshi Kori and Massimiliano Tamborrino for fruitful discussion. 
This study was supported by
the World-leading Innovative Graduate Study Program in Proactive Environmental Studies (WINGS-PES), the University of Tokyo, to T.Y., 
JSPS KAKENHI (Nos. JP22K11919 and JP22H00516) and Japan Science and Technology Agency CREST (No. JPMJCR1913) to H.N., and 
JSPS KAKENHI (Nos. JP18K11560, JP21H03559, JP21H04571, JP22H03695, and JP23K24950) and AMED (Nos. JP21wm0525004 and  JP223fa627001) to R.K.

\begin{figure*}[htb]
  \centering
       \includegraphics[width=17cm]{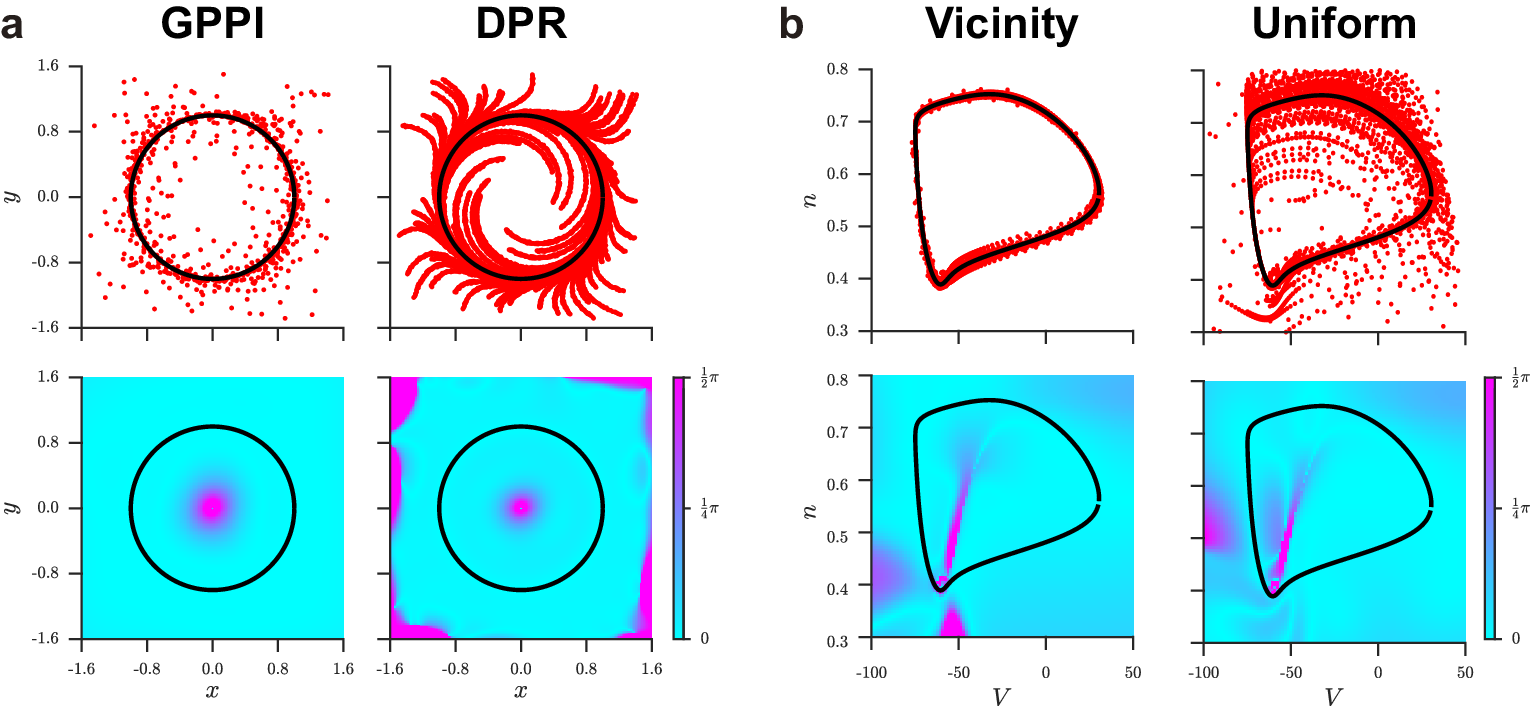}
  \caption{ 
  Estimation error in the phase function. 
  Top panels show the data used for estimating the phase function (red points) and the limit cycle (black). 
  Bottom panels show the estimation error. 
  (a) Stuart-Landau oscillator. (b) Hodgkin-Huxley oscillator.  
}
  \label{fig:Append}
\end{figure*}

\appendix

\section{Derivative phase regression method}
\label{Append:DPR}

We briefly describe the derivative phase regression method (DPR method) proposed by \citet{namura2022estimating}. 
The DPR method estimates two functions, $s(\x):=\sin{\Theta(\x)}$ and $c(\x):=\cos{\Theta(\x)}$, as the proposed method (GPPI method). 
In contrast to the GPPI, which regressed the functions ($s(\x)$ and $c(\x)$) directly using Gaussian processes, 
the DPR regresses the time derivatives of these functions using the polynomial function. 

Here, we describe an overview of the DPR method (see~\citet{namura2022estimating} for details). 
Taking the time derivatives of the the functions ($s(\x)$ and $c(\x)$) using Eq.~\eqref{eq_phasediff}, we obtain
\begin{align}
    \label{eq_DR1}
    \nabla s(\x(t)) \cdot \dot\x(t) &= \omega c(\x(t)), \\
    \label{eq_DR2}
    \nabla c(\x(t)) \cdot \dot\x(t) &= -\omega s(\x(t)), 
\end{align}
where $\nabla$ denotes the gradient of a function, \blue{$\cdot$ denotes the inner product, and $\dot\x(t) := \frac{d}{dt} \x(t)$ denotes the time derivative of $\x(t)$.}
Let us approximate these functions with a polynomial function, 
\begin{align}
    s(\x) &= \sum_{ \blue{k} }{a_kf_k(\x)},\\
    c(\x) &= \sum_{ \blue{k} }{b_kf_k(\x)},
\end{align}
where $f_{\blue{k} }(\x)$ is a polynomial function of the state variables ($x_1, x_2, \cdots ,x_d$). 
The parameters $\{ a_k, b_k\}$ are determined to satisfy Eqs.~\eqref{eq_DR1} and \eqref{eq_DR2}, which is achieved by minimizing the error function $E( \{a_k\}, \{ b_k \} )$, 
\begin{multline}
    E:= \sum_{ \blue{i,j} }{
    \left(\sum_k{a_k\nabla f_k(\x_i\blue{(t_j)})\cdot \dot\x_i\blue{(t_j)} } - \omega\sum_k{b_kf_k(\x_i\blue{(t_j)})} \right)^2
    }\\ + \sum_{\blue{i, j} }{
    \left(\sum_k{ b_k\nabla f_k(\x_i\blue{(t_j)})\cdot\dot\x_i\blue{(t_j)} } + \omega\sum_k{a_k f_k(\x_i\blue{(t_j)})} \right)^2
    },
\end{multline}
\blue{where the sum is taken over the time series $i= 1, 2, \cdots, n$ and the time index $j$.} 
We also require a constraint to fix the origin of the estimated phase function $( \Theta(\x_{\blue{\rm orig} })= 0 )$, i.e., 
\begin{align}
    \label{eq_Const1}
    s(\x_{\blue{\rm orig} }) &= \sum_{\blue{k} } {a_k f_k(\x_{\blue{\rm orig} })} = 0,\\
    \label{eq_Const2}
    c(\x_{\blue{\rm orig} }) &= \sum_{\blue{k} } {b_k f_k(\x_{\blue{\rm orig} })} = 1.
\end{align}
The minimization problem of the error function $E( \{a_k\}, \{ b_k \} )$ with the constraint (Eqs. \ref{eq_Const1} and \ref{eq_Const2}) is a quadratic programming. Thus, we can determine the parameters $\{ a_k, b_k\}$ using an efficient solver, such as quadprog in MATLAB. 
Finally, the phase function $\Theta(\x)$ is obtained from the estimate of the functions ($s(\x), c(\x)$) using Eq.~\eqref{eq_arctan}.

\section{Estimation error of the phase function} \label{data_error}
We evaluate the estimation performance of the asymptotic phase function by calculating the error: $E(\x)= |\hat{\Theta}(\x) - \Theta(\x)|$, where $\hat{\Theta}(\x)$ and $\Theta(\x)$ represent the estimated and true phase functions, respectively.  
Figure \ref{fig:Append}a shows the data used for estimation and the estimation error in the phase function for the Stuart-Landau oscillator, which corresponds to Fig.~\ref{fig:SLphase} (Sec.~\ref{sec:SL_model}). While the DPR method used all the 50,000 data points for estimating the phase function, the proposed method (GPPI) used only 1000 data points sampled from the data. The strength of the observation noise was $\eta=0.005$. 
Figure \ref{fig:Append}b shows the data used for estimation and the estimation error in the phase function for the Hodgkin-Huxley oscillator, which corresponds to Fig.~\ref{fig:HHphase} (Sec.~\ref{sec:HH_model}).

\bibliographystyle{elsarticle-harv}
\bibliography{bib_arxiv}

\end{document}